Eric E. Hamke, Ramiro Jordan, Manel Ramon-Martinez

# Breath Activity Detection Algorithm

***Abstract*— This report describes the use of a support vector machines with a novel kernel, to determine the breathing rate and inhalation duration of a fire fighter wearing a Self-Contained Breathing Apparatus. With this information, an incident commander can monitor the firemen in his command for exhaustion and ensure timely rotation of personnel to ensure overall fire fighter safety.**

***Index Terms* - ???**

## I. INTRODUCTION

The Enhanced Situational Awareness in Firefighting Emergencies project seeks to expand the knowledge available about the incident in an integrated manner and make this data available to firefighting incident commanders and other firefighters working the incident.

### A. Firefighting Data Collection Systems

The intention is to use the existing communication systems with minimal changes to the certified and approved systems currently in use. Currently, the primary tools available for communication and data collection are analog or digital radios. Ultrahigh frequency (UHF) radios are used in a network where the radio frequency is the shared resource. The prevalence of these systems is a reliance on low cost and low complexity that guarantees the systems robustness and availability. Another advantage is that firefighters can join or dropout as needed resulting in a very flexible network that allows for a high degree of autonomy for the individual firefighters. Thus any system designed to operate in this environment must rely on local processing of data related to each firefighter and communicate only pertinent information without interfering with voice communications.

The disadvantages to using a system like this are that the voice communications reflect only what the firefighter wishes to communicate about his perceptions of the situations he is in. The data may be incomplete, since the firefighter's focus is localized to the immediate tasks at hand. Further, because of the focus, he may forget information since it is not part of the immediate task. Additionally, there is information being recorded by the microphones that is normally suppressed to make the communications clearer.

This paper presents the beginnings of the process of extracting additional data using the radio microphones in conjunction with a small signal processing unit. From the additional audio data, it is the hope that we will be able to detect background noises, related to explosions, various alarms that the fireman wears to warn of situations such as carbon monoxide and hydrogen cyanide released by burning plastics, Self-Contained Breathing Apparatus' low air alarm and the firefighter stationary or inactive sensors intended to alert other firefighters when a firefighter is injured or overcome.

In addition, we can also monitor the regulator sounds of the Self-Contained Breathing Apparatus (SCBA) to detect the starting and ending time of the inhalation events. These times can then be processed to compute the breathing rate and inhalation duration. Further processing could compute the accumulated inhalation durations to estimate the amount of air remaining in the fire fighter's air tank. This information and the breathing rate combined with other physiological data could be relayed back to the incident commander at regular intervals using burst communication and can be encoded into the voice codex used by the radios.

### B. Previous Works

Others have studied the process of monitoring breathing sounds with the intent to identify them as noise that needs to be eliminated from a recording or speech processing system. The breath sound can be mistaken by speech processing systems as a fricative sound and result in a false positive detection of certain phonemes. Several medical applications have focused in determining breathing rates of patients on respirators [16]. The work using intrusive devices such as a respiratory inductive plethysmography net. The data from the strain gauges is then processed using an Artificial Neural Network. Another medical study involved segmenting simulated breathing cycles. The process proposed by Li et.al, [17] involves using a gradient to determine the start of breathing cycles in a tidal breathing pattern. The second stage of the process is to examine the time between starting cycles and their duration. Since a breath is a long duration compared to phonemes and can last for 1 second or more.

Another area of research involves speech related applications. The intention in these applications is to identify breathing artifacts and remove them from the speech. The primary features used are based on cepstral or meI frequency cepstral coefficients (MFCC). Price et al. automatic breath detection methods of speech signals were based on cepstral coefficients and used Gaussian Mixture Model (GMM) as the classifier, and achieved a detection rate of 93% [18], Wightman and Ostendorf algorithm uses Bayesian classifier and achieved

Eric. Hamke, Department of Electrical and Computer Engineering, University of New Mexico, Albuquerque, NM 87110 USA (e-mail: ehamke@unm.edu).

a detection rate of 91.3% [19]. Ruinskiy and Lavner applied algorithm of automatic breath detection in speech and songs based on MFCC parameters, short-time energy, ZCR (zero-crossing rate), spectral slope, and duration [1], They introduced also a method of edge detection, They achieved precision/recall rate 97,6%/95.7%. Nakano et al. described detailed acoustic analysis of breath sounds using MFCC parameters and Hidden Markov Models [20]. The result was 97.5%/77.7% rates for unaccompanied singing voice.

There are also many works performing breath detection as one of many audio events in audio retrieval systems. Most of them use MFCC or LPC (Linear Predictive Coding) parameters [21]. In Classification the extracted feature vectors are assigned to one of a set of classes. In MARSYAS, the Gaussian (MAP), Gaussian Mixture Model (GMM) and K−NN families of statistical Pattern Recognition classifiers are supported. Two case studies of classification have been implemented and evaluated in our system: a music/speech classifier and a genre classifier. The music/speech classifier achieves a 90.1% classification accuracy.

M. Igras, B. Ziolko [23]propose a method involving both temporal and spectral features. The time domain analysis of the extracted breaths determines the length and energy (normalized in reference to the entire speech signal) is computed. These features are combined with a discrete wavelet analysis. Finally the wavelet features and the time features are classified using a dynamic time warping algorithm, where the time-axis fluctuation is approximately modeled with a nonlinear warping function .Igra et. al., reported a 94.7% success at detecting breaths with this approach.

S. Borys and M. Hasegawa-Johnson investigated several techniques including Hidden Markov Models and Support Vector Machines (SVM). They trained groups of SVM classifiers that could detect different landmarks. The first group of trained classifiers were designed to distinguish between transitions for the features speech, sonorant, continuant and syllabic. The second group of SVM classifiers was trained to recognize positive and negative gradients where feature could be speech, consonantal, sonorant, +consonantal/continuant. It is not clear which types of kernels tey used. The kernels alluded to include linear, radial basis function, polynomials and sigmoid functions. Optimization for each different kernel follows the same general procedure as the optimization for the linear kernel. They concluded in their study that the SVMs they developed had approximately 85% for Syllabic features and 97% for speech. The remaining features are in 95% for sonorant and 90% for continuant.

The organization of the presentation starts by briefly discussing the aspects of the SCBA for those who are not familiar with the device. The next section is a survey of the two methodologies reviewed, a pattern matching and the use of a Linear Predictive Coefficients based filter. Additionally, the third section will deal with the process of making the exemplars used in the process. The feature extraction/enhancement process is presented in the next section. In the detection section we propose the use of machine learning (support vector machines with a novel kernel). The final sections will deal with the detection process.

## II. ACOUSTIC PROPERTIES AND EFFECTS OF SELF-CONTAINED BREATHING APPARATUS

The investigations focus on monitoring the breathing rate of the firefighter using a Self-Contained Breathing Apparatus (SCBA). The SCBA system is a pressure-demand air delivery system. When a user inhales negative pressure within the mask causes the regulator valve to open. Pressurized air enters the mask producing a loud broadband hissing noise. The noise is comparable in amplitude to the speech signal (fricative sound). This signal is broadband and incoherent.

The mask (Fig. 1) is a rigid structure with a clear plastic face plate and a flexible rubber seal that contacts the forehead temples cheeks and chin of the wearer. The SCBA systemnoises includes low air alarms and air regulator noises as well. [13]

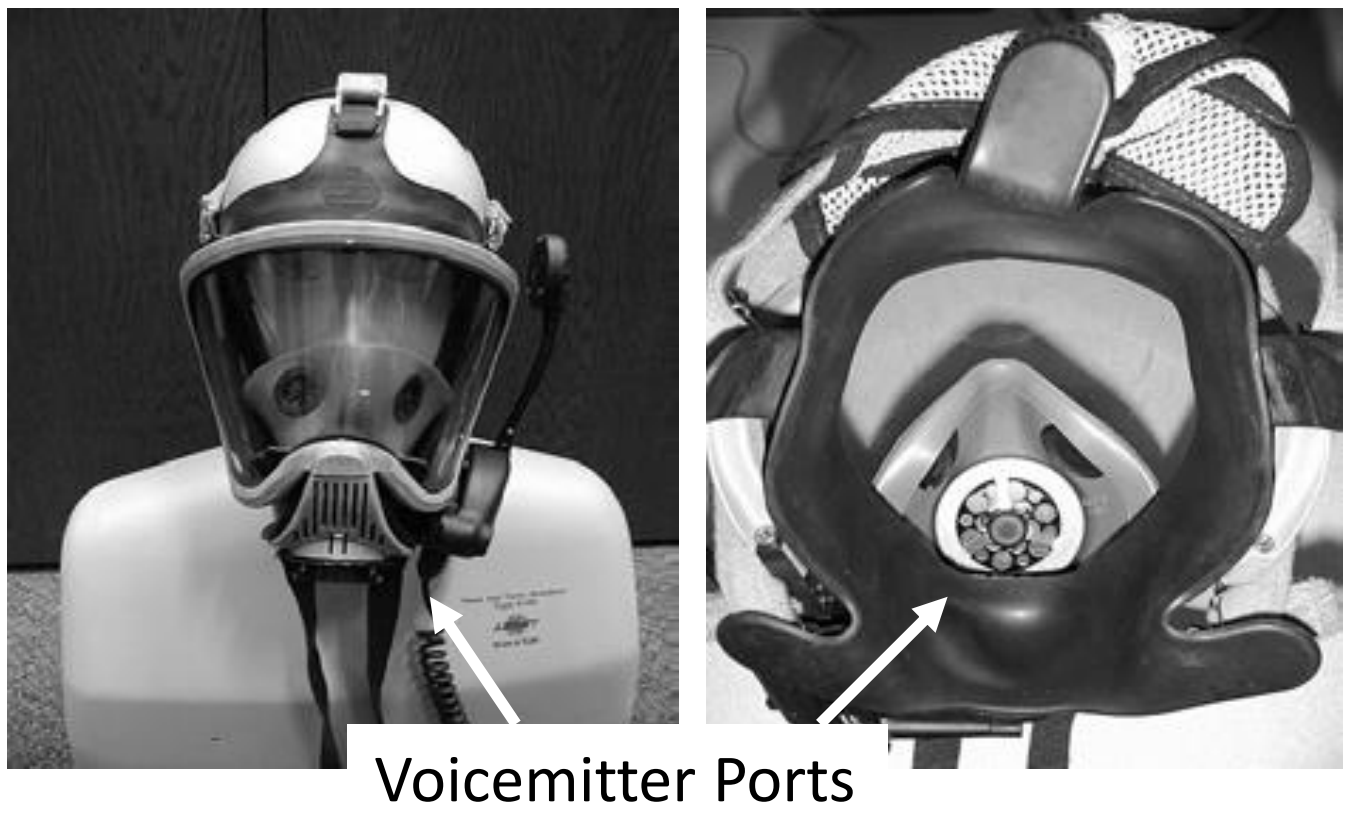

Fig. 1. External and internal views of a commonly used SCBA mask showing the voicemitter port. The breathing cycle is divided into four different phases: [13]

## III. THE METHODOLOGIES

The movement of air in from the tank makes a very distinctive sound (Fig. 2) that resembles a fricative sound. Fricative sounds are the part of speech used for forming consonants such as "f". The detection of these sounds along with other aspects of taking a breath (Fig. 3) have been the subject of several papers.[1], [2], [12] Their intended use is in the music and entertainment industry signal processing where these sounds are remove or enhance for clarity or artistic reasons. The approaches discussed are to identify the inspiratory or inhalation phase since the expiratory phase is generally used to generate the formants of speech.

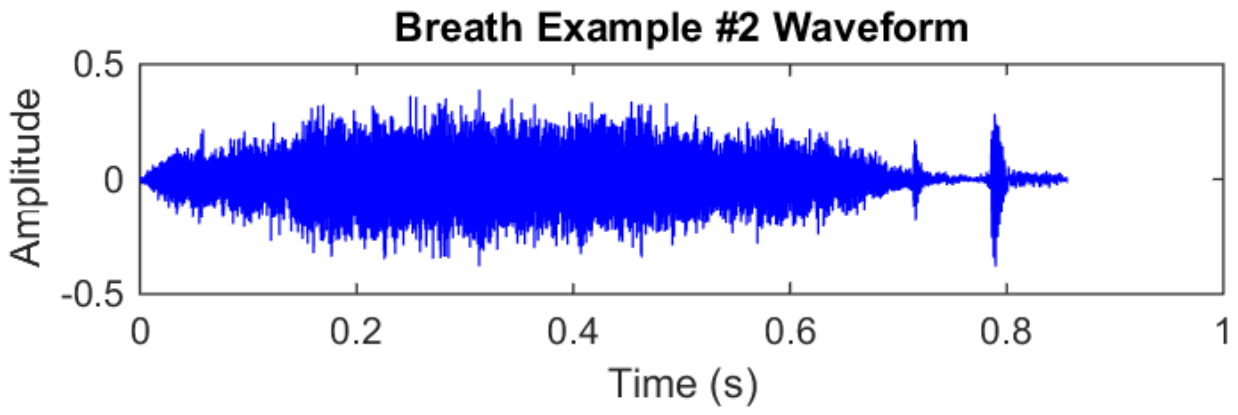

Fig. 2, The inspiratory phase is distinctly present and the clicking sound of the regulator's valves as they reset for the expiratory phase are shown at the end of the sound bite as2 sharp energy spikes.

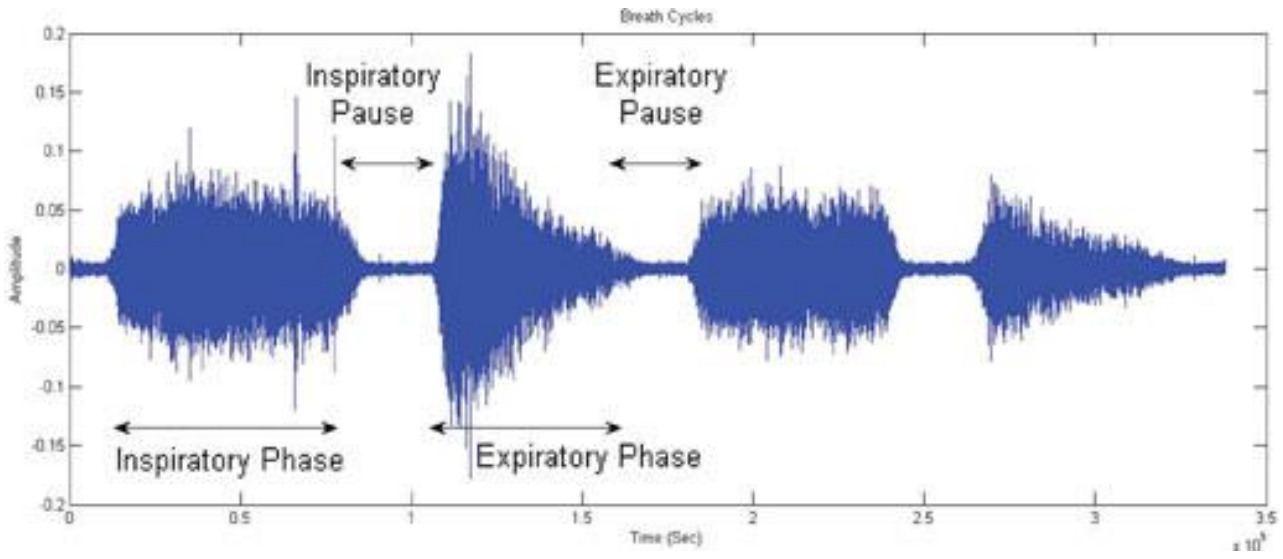

Fig. 3. The breathing cycle is divided into four different phases:

1. inspiratory phase - moment the air inflow starts. When the airflow stops, the inspiratory phase ends;
2. inspiratory pause - pause begins and lasts until the air begins to flow out from the lungs;
3. expiratory phase - has a higher average energy than the inhale, which is due to the fact that air exits the lungs in addition to the voice that is generated
4. expiratory pause - lasts until the end of the breathing cycle.

Acoustic Signal Classification of Breathing Movements to Virtually Aid Breath Regulation [12]

### A. *Pattern Matching*

Ruinskiy and Lavner [1] propose automating breath detection by processing the short time (windowed) cepstrum using an image recognition technique. The generalized approach is shown in Fig. 4. The cepstrum is computed for the sound being filtered using a mel frequency scale. The mel scale relates perceived frequency, or pitch, of a pure tone to its actual measured frequency in an attempt to match how people hear sounds.

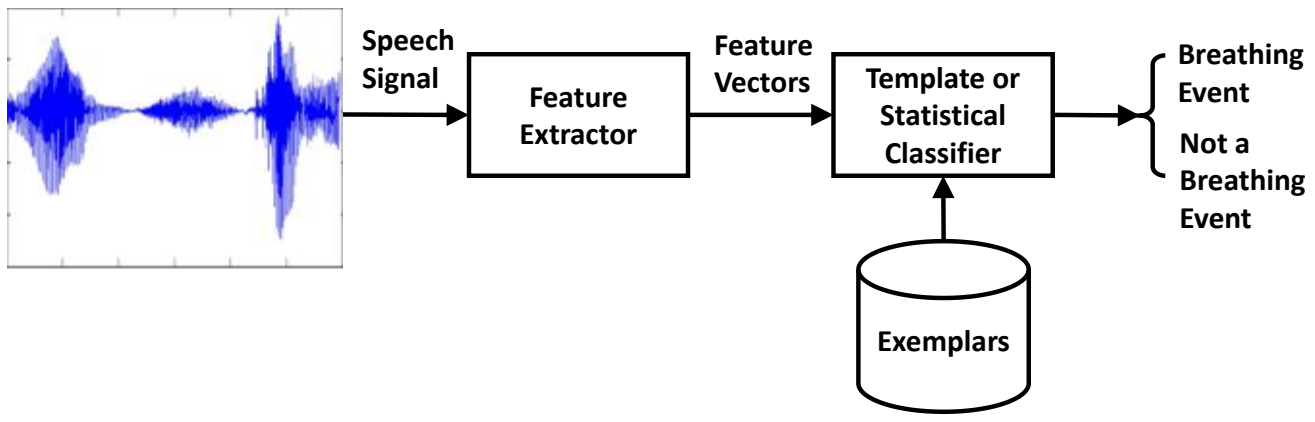

Fig. 4, Data flow for automatic breath detection using Ruinskiy and Lavner.

### B. *Linear Predictive Coefficients*

The ARINA algorithm is initialized using a set of autocorrelation coefficients derived offline from a sample of the inhalation noise. A 10th order Linear Predictive Coefficients (LPC) all pole filter. The filter is then inverted into a Finitie Impulse Response (FIR) filter and used to filter the speech signal of the fireman. Kushner et. al. also proposed using a moving average to update the autocorrelation coefficients and recomputing the LPC continuously to adapt to different wearers and different masks.

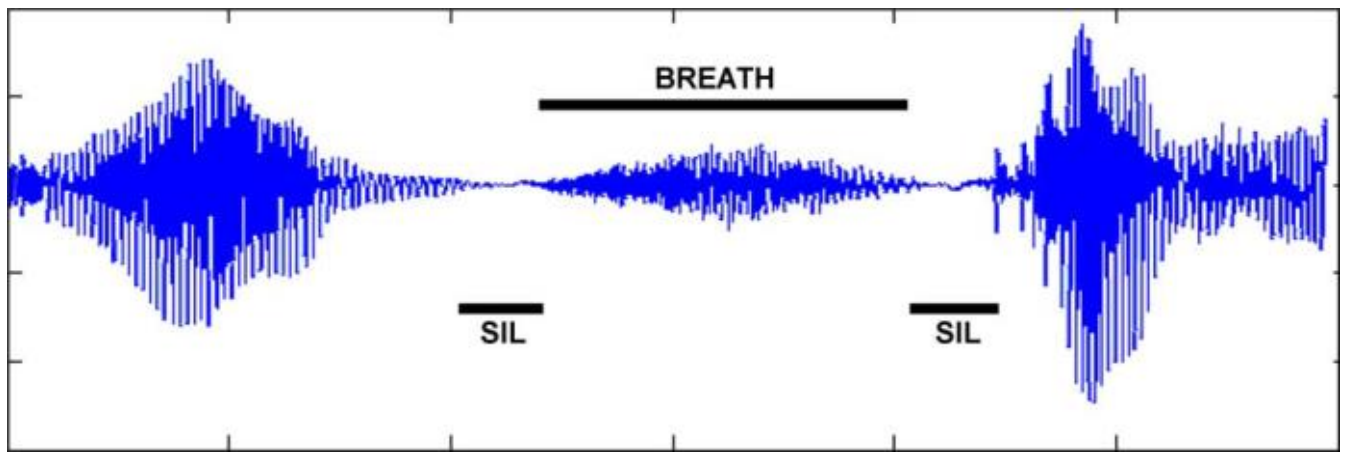

Fig. 5, Part of a voice waveform demonstrating a breath sound located between two voiced phonemes. The upper line marks the breath, characterized by higher energy near the middle and lower energy at the edges. The lower lines denote the silence periods (SIL) separating the breath from the neighboring phonemes

An Effective Algorithm for Automatic Detection and Exact Demarcation of Breath Sounds in Speech and Song Signals [1]

### C. *Creating Exemplars*

Regardless of the approach used, it is necessary to identify a set of examples or exemplars to be used for training. The process of building exemplars involves isolating multiple instances of the air rushing into the mask in the recorded sound track. These sounds usually are corrupted with the low air alarm sound which is a clacking noise at approximately 28Hz (Fig. 6). The regulator inspirational sound with the low air alarm superimposed was used to form the exemplars.

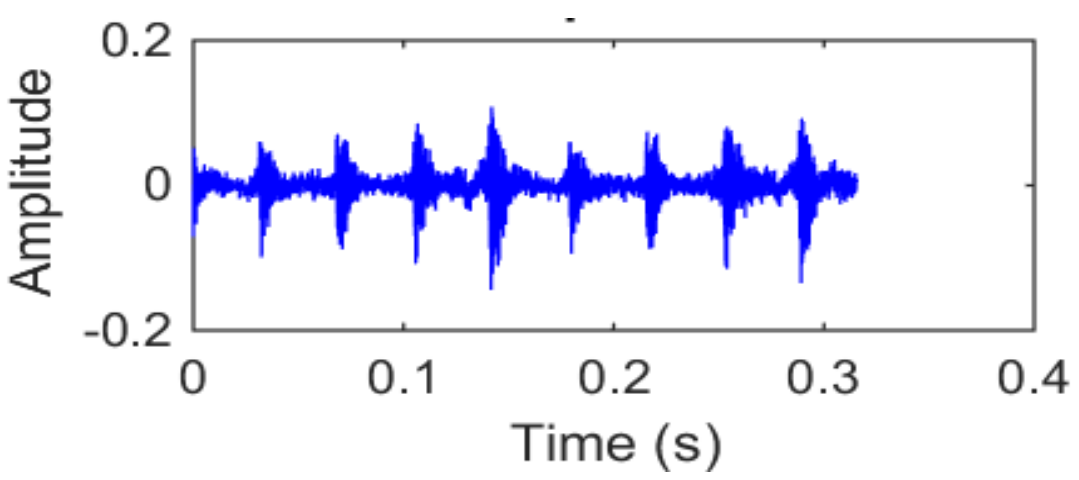

Fig. 6, The low air alarm is a cyclic sound made with a clacker at approximately 28Hz,

Using a sound mixing tool, the exemplars of the sounds are isolated and recorded into wave files. Each exemplar recording is sectioned into 15 ms frames with an overlap 5 ms. The frames are then windowed with a half-hamming window to smooth edges and de-emphasize high frequency components. The windowed frames are pre-emphasized using a first-order difference filter.

In our implementation of the LPC approach, we used a shortened exemplar to see if we could develop a measurement of signal duration as well as measure the time between inhalations. The shortened exemplars approach did not work as well for the pattern matching approach. Since the similarity between consonants, thermal noise and the inhalation noise is too close to distinguish between the three. The LPC approach's real strength is the ability to make this distinction,

## IV. Feature Extraction/Enhancement

The general structure of the algorithms is shown in Fig. 7. The pre-emphasis stage usually involves the removal or minimization of non-speech sound. The windowing frames the signal into short time intervals (20 to 40ms wide).

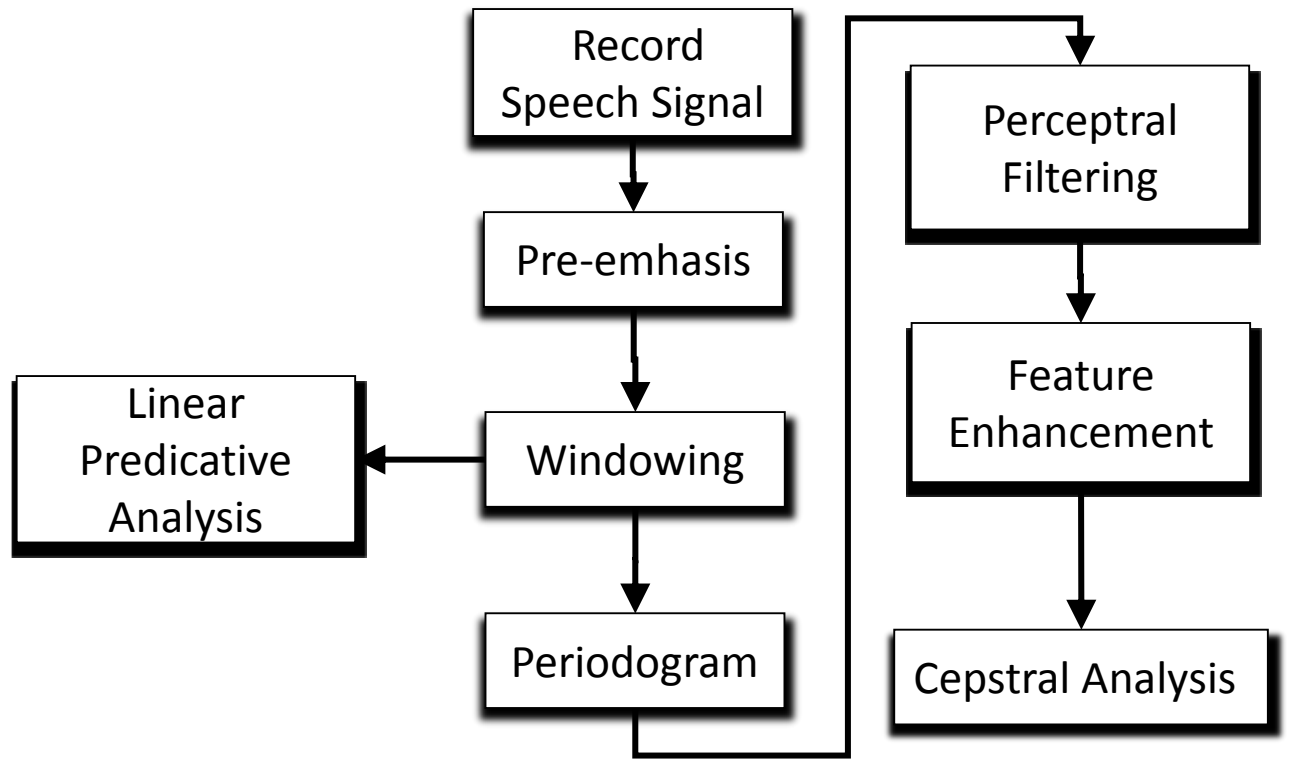

Fig. 7, General data flow for breath detection algorithms reviewed

The windowing step is done for three reasons. First since we are looking to develop a near real time algorithm we wish to have the data in as small a window as possible that still allows for reliable signal processing results. The framing also allows for the assumption of statistical stationary for the short time scales. (If the frame is too short there is not enough samples to get a reliable spectral estimate; if it is longer the signal changes too much throughout the frame.) A half Hamming window is composed by changing the lower frequency elements to a weighting of one and leaving the other window weights unchanged. These changes de-emphasize the higher frequencies while keeping the lower frequencies intact.

### A. Feature Extraction using MEL Frequency Coefficients

Ruinskiy and Lavner [1] begin their approach with a periodogram to estimate the squared spectral density of the windowed speech signal.

$$S_i[k] = \left| \sum_{n=1}^{N} s_i[n] h[n] e^{-j2\pi kn/N} \right|^2, \quad 1 \le k \le K \qquad (1)$$

where $h[n]$ is an $N$ sample long analysis window (e.g. Hamming window), and $k$ is the length of the Discrete Fourier Transform (DFT). The periodogram-based power spectral estimate for the speech frame $S_i[n]$ is given by:

$$P_i[k] = \frac{1}{N} \left| S_i[k] \right|^2 \qquad (2)$$

This is motivated by the human cochlea (an organ in the ear) which vibrates at different spots depending on the frequency of the incoming sounds. Depending on the location in the cochlea that vibrates (which wobbles small hairs), different nerves fire informing the brain that certain frequencies are present. Our periodogram estimate performs a similar job for us, identifying which frequencies are present in the frame. [2]

The Perceptual filtering stage in processing the windowed signal is to use a perceptually based filterbank. The papers reviewed ([1], [2]) generally used on of two types of filterbanks, the BARK and MEL scales. The BARK scale [3] was not used in this investigation but represents another possibility that could be explored in future work.

The MEL scale relates perceived frequency, or pitch, of a pure tone to its actual measured frequency. Humans are much better at discerning small changes in pitch at low frequencies than they are at high frequencies. Incorporating this scale makes our features match more closely what humans hear. As a reference point, the pitch of a 1 kHz tone, 40 dB above the perceptual hearing threshold, is defined as 1000 mels. The formula for converting from frequency to Mel scale is [4]

$$M(f) = 1125 ln(1 + f / 700) \qquad (3)$$

A filter bank constructed using the mel scale would appear as in Fig. 8 [4]. The filter equation is

$$H_m[k] = \begin{cases} 0 & k < f(m-1) \\ \dfrac{k - f(m-1)}{f(m) - f(m-1)} & f(m-1) \le k \le f(m) \\ \dfrac{f(m+1) - k}{f(m+1) - f(m)} & f(m) \le k \le f(m+1) \\ 0 & k > f(m+1) \end{cases} \qquad (4)$$

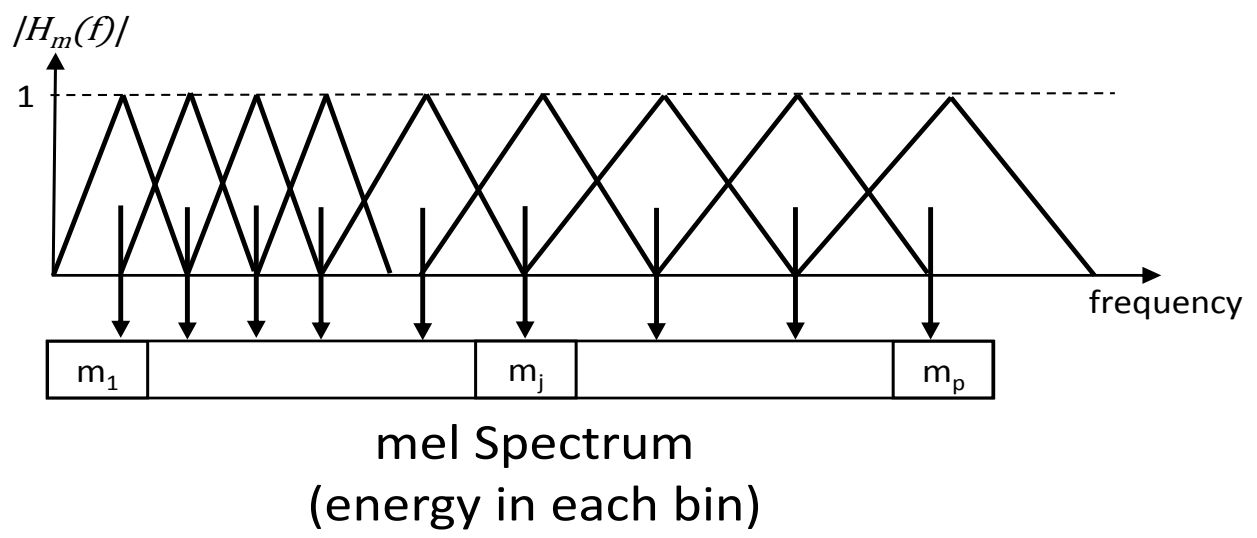

Fig. 8, The first filter starts at the first point, reach its peak at the second point, then return to zero at the 3rd point. The second filter will start at the 2nd point, reach its max at the 3rd, then be zero at the 4th etc. [4]

An pattern matching approach a Discrete Cosine Transform (DCT) is used to eliminate redundant information from the data by eliminating the correlations between data points.[1] The transform expresses a finite sequence of data points in terms of a sum of cosine functions oscillating at different frequencies.

$$c_i = \sqrt{\frac{2}{N}} \sum_{j=1}^{N} m_j cos\left( \frac{\pi i}{N} (j - 0.5) \right) \qquad (5)$$

where $N$ is the number of filter bank channels and $m_j$ are sequence of logarithms of the MEL frequency periodogram.

Cepstrum analysis is a nonlinear signal processing technique to determine the fundamental frequency of human speech. The computation of the cepstral coefficients is done as follows.

$$\underline{\hat{x}} = \text{IFFT}\{\ln(|\text{FFT}\{f_n\}|)\} \qquad (6)$$

Cepstrum pitch determination is particularly effective because the effects of the vocal excitation (pitch) and vocal tract (formants) are additive in the logarithm of the power spectrum and thus clearly separate. [2] The concept behind cepstral analysis is deconvolution. In the speech research field, the main use of cepstral analysis is to extract the vocal tract characteristics from a speech spectrum. It can be shown that a speech sequence can be modelled by the convolution of the vocal tract impulse response and the glottal excitation.

$$s[n] = v[n] * u[n] \Leftrightarrow \tilde{s}[n] = \tilde{v}[n] + \tilde{u}[n] \qquad (7)$$

where $v[n]$ is the impulse response of the vocal tract filter; $u[n]$ is the glottal excitation, which is usually a quasi-periodic impulse sequence for voiced speech; $\tilde{s}[n]$ is the cepstrum of the speech sequence; $\tilde{v}[n]$ is the cepstrum of the vocal tract impulse response sequence; $\tilde{u}[n]$ is the cepstrum of the glottal excitation sequence. [2]

Finally a liftering operation is performed to further isolate and identify the coefficients of interest. The principal advantage of cepstral coefficients is that they are generally decorrelated and this allows diagonal covariances to be used in the hidden Markov models (HMMs). However, one minor problem with them is that the higher order cepstra are numerically quite small and this results in a very wide range of variances when going from the low to high cepstral coefficients, $c_n$. The value $L$ to lifter the cepstra according to the following formula [5]

$$c_n = \left(1 + \frac{L}{2} sin\left(\frac{\pi n}{L}\right)\right) c_n \tag{8}$$

Next a matrix is formed by concatenating MFCC vectors for each subframe. Each such matrix is denoted, by $\underline{\underline{M}}_i$ where $i = 1, \ldots, N$ and $N$ is the number of exemplars used to build the template.

The DC component is removed by subtracting the mean for each subframe MFCC vector. An estimate of the unbiased mean cepstrogram is found by averaging the matrices across the sample sets:

$$\underline{\underline{T}} = \frac{1}{N} \sum_{i=1}^{N} \underline{\underline{M}}_i \tag{9}$$

The variance is estimated using the sample standard deviations for the cepstrograms:

$$\underline{\underline{V}} = \sqrt{\frac{1}{N-1} \sum_{i=1}^{N} \left(\underline{\underline{M}}_i - \underline{\underline{T}}\right)^2} \tag{10}$$

The resulting templates are shown in Fig. 9.

A singular value decomposition (SVD) of the resulting matrix is computed on the Unbiased MFCCs Template. The First singular vector ( $\underline{S}_1$ ) corresponding to the largest singular value is derived. This vector is expected to capture the most important features of the breath event, and thus, improve the separation ability of the algorithm when used together with the template matrix in the calculation of the breath similarity measure of test signals.

### *B. Feature Extraction using LPC*

In human speech production, the shape of the vocal tract governs the nature of the sound being produced. In order to study the properties quantitatively, the vocal tract is modelled by a digital all-pole filter. [6], [2] The following equation expresses the transfer function of the filter model in z-domain,

$$V(z) = A \frac{1}{\sum_{k=1}^{N} a_k z^{-1}} \tag{11}$$

where $V(z)$ is the vocal tract transfer function. $G$ is the gain of the filter and { $a_k$ } is a set of autoregression coefficients called Linear Prediction Coefficients. The upper limit of summation, N, is the order of the all-pole filter. The set of LPC determines the characteristics of the vocal tract transfer function.

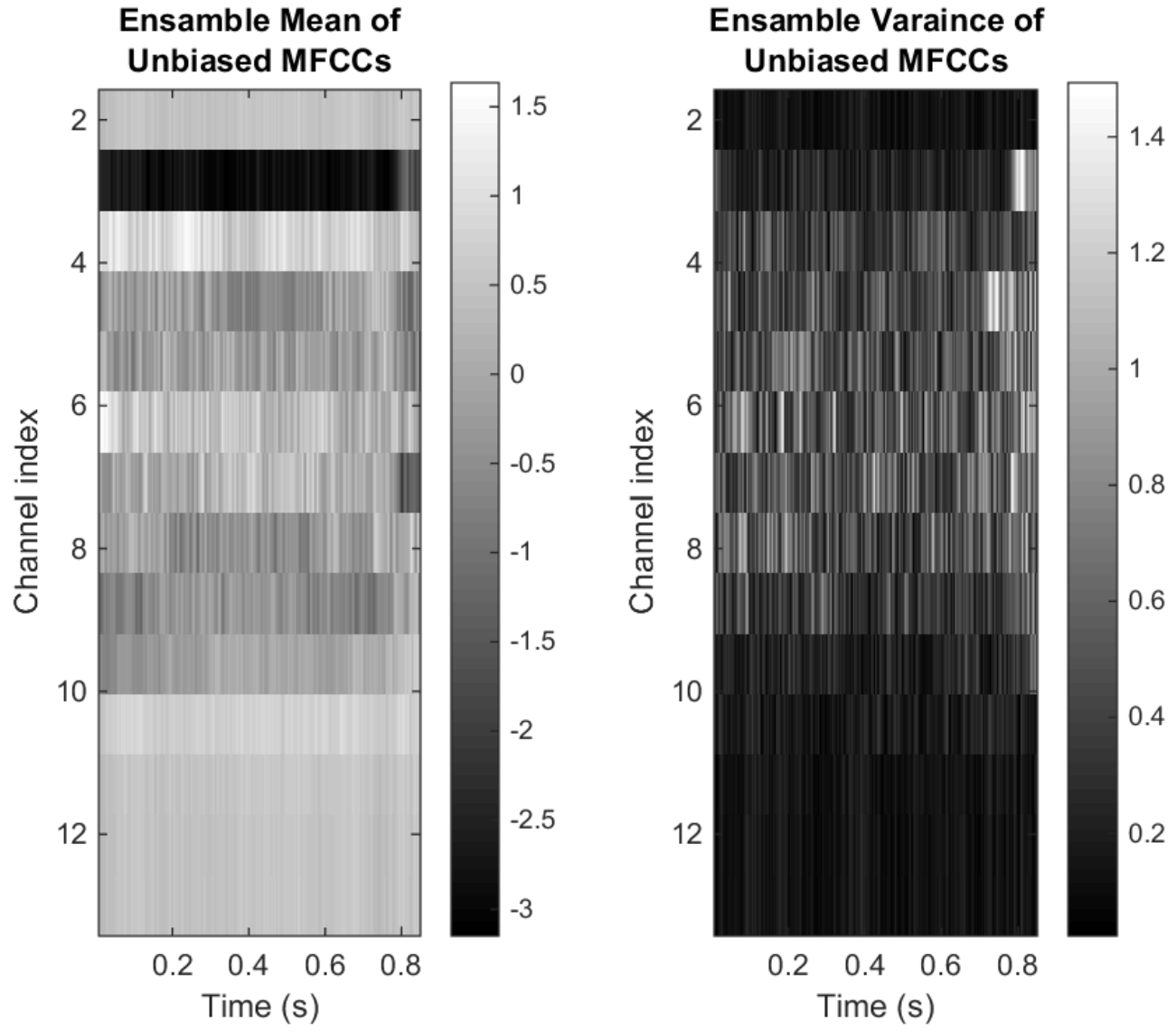

Fig. 9, The resulting mean and variance Templates for the exemplars.

Solving for the LPC involves calculating a matrix of simultaneous equations $(R_x)$ and the autocorrelation of the windowed speech frames $(R_{xx}[n])$.

$$R_x = E\left\{x[n]x^H[n]\right\} = \begin{bmatrix} R_{xx}[0] & \cdots & R^*_{XX}[N-1] \\ \vdots & \ddots & \vdots \\ R_{xx}[N-1] & \cdots & R_{xx}[0] \end{bmatrix} \tag{12}$$

The following equation is then solved for the autoregression coefficients.

$$R_x \begin{bmatrix} a_1 \\ \vdots \\ a_N \end{bmatrix} = \begin{bmatrix} R_{xx}[1] \\ \vdots \\ R_{xx}[N] \end{bmatrix} \tag{13}$$

The gain of the all-pole filter can be found by solving the following equation.

$$A = \sqrt{R_{xx}[0] - \sum_{k=1}^{N} a_k R_{xx}[k]} \tag{14}$$

## V. Detection Criteria

### *A. Pattern Matching Approach*

In the study, the entire file is read in and processed into a cepstrogram. A sliding window is then passed over the data. The window is slide to the right by one cepstral column and the process of computing the breathiness index is performed using the cepstogram for the data falling inside the window.

The normalized the difference matrix is computed first.

$$\underline{\underline{D}} = \left(M\left(\underline{\underline{X}}_i\right) - \underline{\underline{T}}\right) / \underline{\underline{V}} \tag{15}$$

The normalization by the variance matrix ($V$) is performed element-by-element. Normalizing compensates for differences in the distributions of the various cepstral columns.

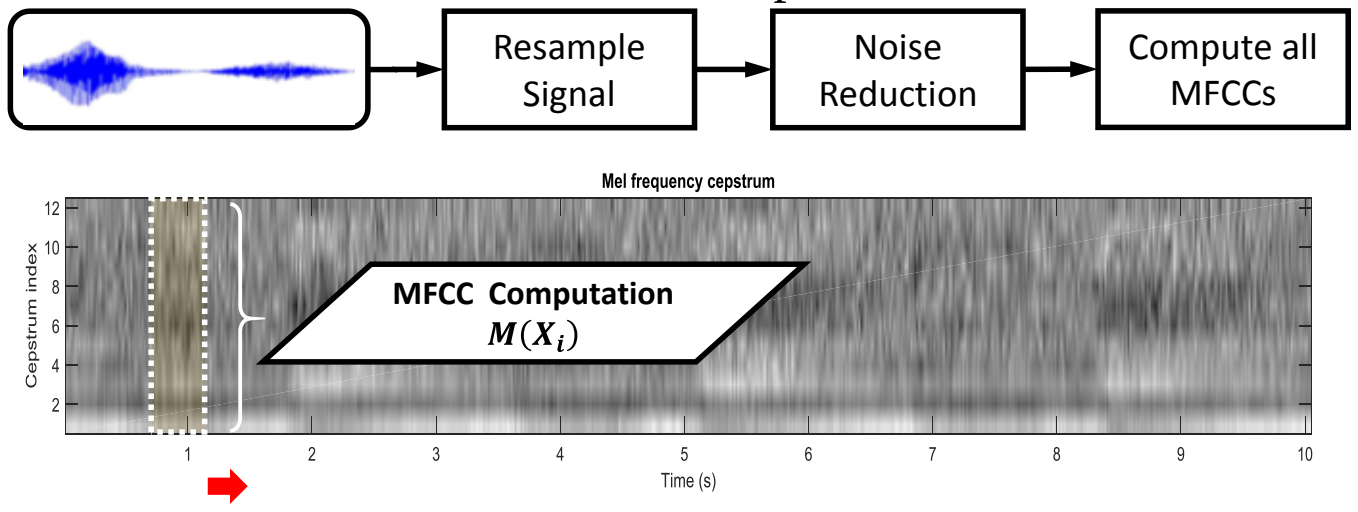

Fig. 10, **P**rocessing of the recording using sliding window.

A lifter is use on each column with a half-Hamming window to emphasize the lower cepstral coefficients. This liftering improves the separation between breath sounds and other sounds.

The first similarity measure, is then computed

$$C_p = \left( \sum_{i=1}^{n} \sum_{j-1}^{N} \left[ \underline{D}_{ij} \right]^2 \right)^{-1} \tag{16}$$

where $D_{ij}$ are the elements of the normalized difference matrix, $n$ is the number of subframes, and $N$ is the number of MFCC coefficients computed for each subframe.

The second similarity measure is computed using the SVD vector.

$$C_n = \left( \sum_{j=1}^{N} \underline{S}_i, \underline{D}_j \right)^{-1} \tag{17}$$

where $\underline{S}_1$ is the singular vector and $\underline{D}_j$ are the normalized columns of the cepstrogram.

The breath similarity measure,

$$B\left(\underline{X}_i, \underline{T}, \underline{V}, \underline{S}_1\right) = C_p C_n \tag{18}$$

These measures are then categorized based on the peaks (greatest similarity between the template and the windowed signal). The peak finding process locates the local maxima (peaks) of the input signal vector, data. A local peak is a data sample that is either larger than its two neighboring samples or is equal to Inf (infinity). Non-Infinite signal endpoints are excluded.

Another method uses a threshold technique to measure the duration of a peak (Fig. 11) and allows for an estimate of the breath duration.

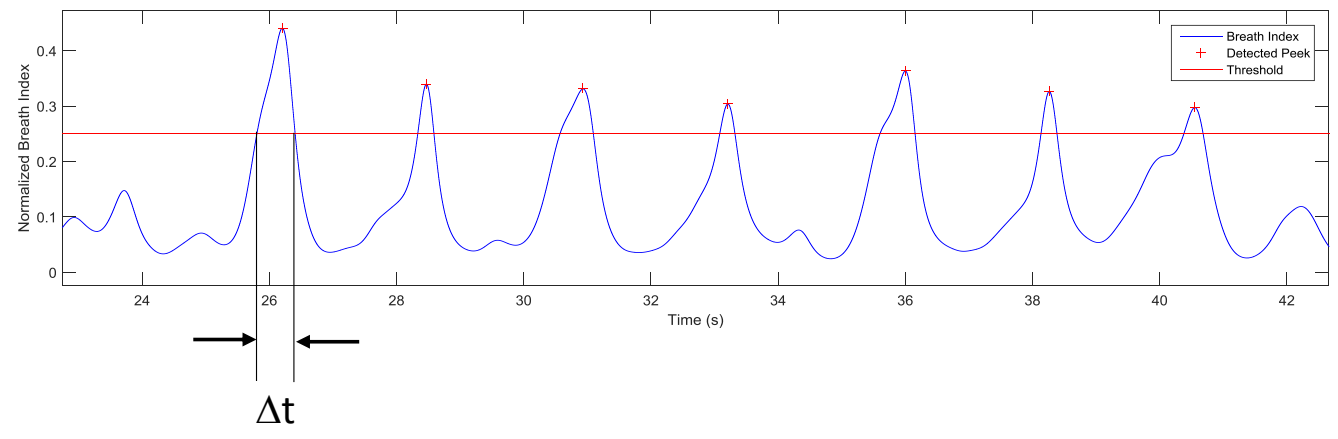

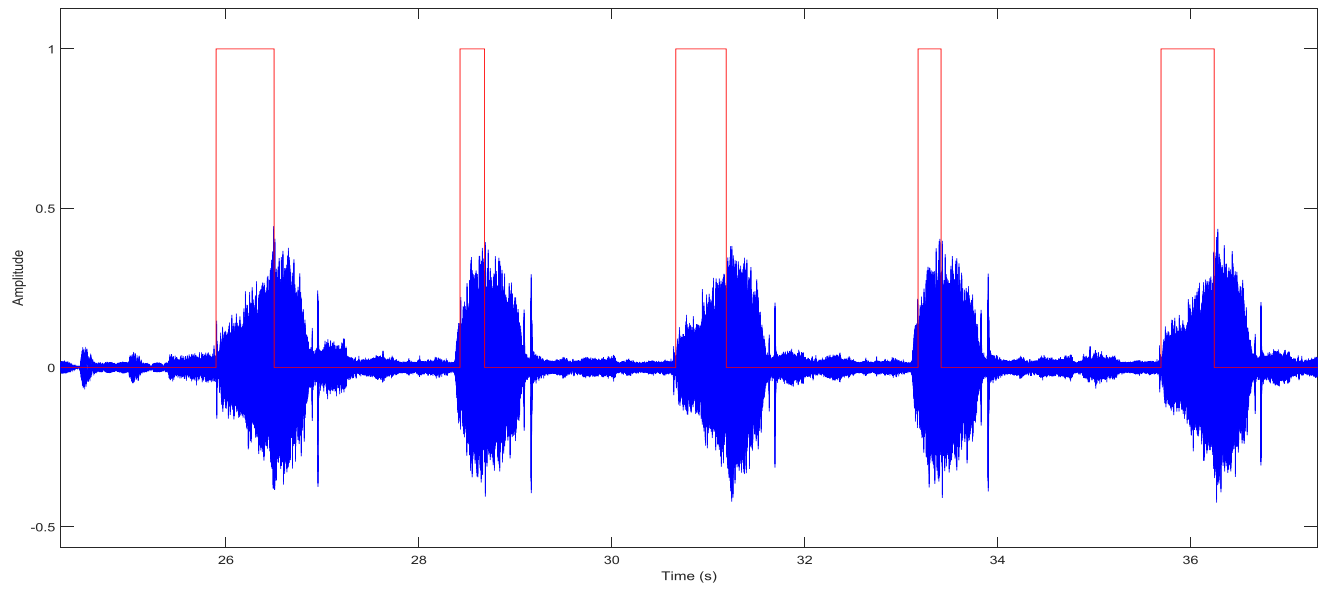

Fig. 11, **M**easuring peak duration.

The duration of the inhalation noise is fairly long compared to unvoiced speech. So we can use a duration threshold test to eliminate any false detection due to speech. Thus the threshold must be met for N consecutive frames before detection is validated.

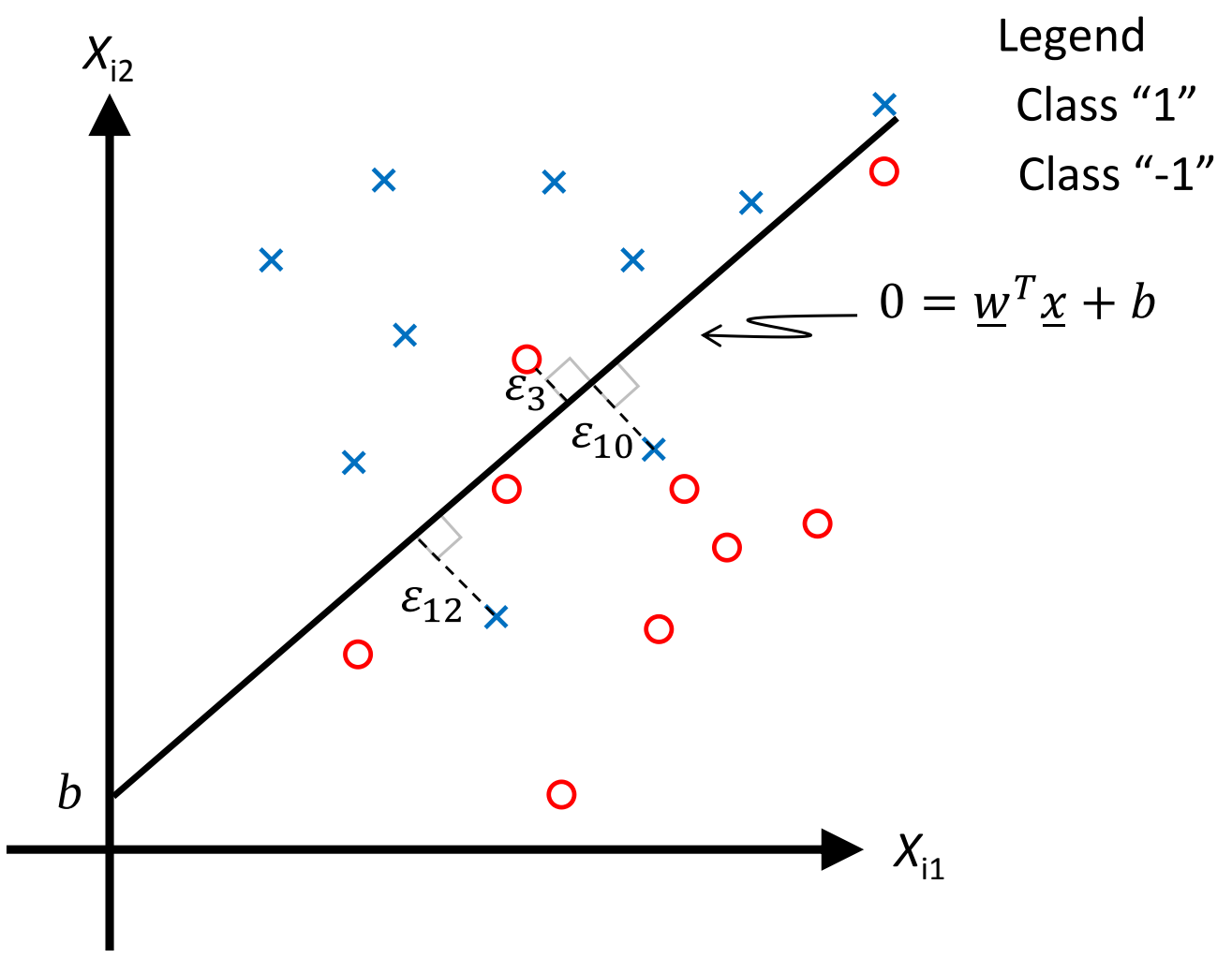

Fig. 12 Binary classification of data in a two-dimensional plane.

### *B. LPC Approach*

Once the coefficients have been determined by fitting the LPC filter to an exemplar. The filter is inverted so that we have a Finite Impulse Response inversion filter $\Lambda(z)$.

$$\Lambda(z) = \sum_{k=1}^{N} a_k z^{-1} \tag{19}$$

The inversion filter then generates an estimate of the input signal. Ideally, if the input to the filter is white noise then we should get the formant or fricative sound being modelled. In this application, the gain ($G_{frame}$) of the estimated input to the actual measured output is computed.

$$G_{frame} = \left( \frac{rms\left(x_{est}[n]\right)}{rms\left(y[n]\right)} \right)^2 \Bigg|_{frame} \tag{20}$$

Like the breath indices before, the values can be thresholded to determine both breathing interval and breath duration.

In addition to this, it is also necessary to check inhalation durations. Some durations are too short like a spike for 2 or 10 frames (2 ms to 20 ms ). Other durations that are too long (e.g.

1 minute or more). The method also may classify durations with minimal power compared to the power level usually present on the in the regulator inflow.

### C. Support Vector Approach

A support vector machine seeks to identify a set of vectors that span the data space. The resulting vector basis is used to describe each data point as a linear combination of the basis vectors. Intuitively, the basis vectors can be thought of as a set of coordinate axes and the linear combination values as the coordinates in that space. It is hoped that by finding the coordinate system, we can use it to draw a hyper-plane that separates the data into groups or classifications. Vapnik [14] theorized that the problem of training a system to classify a set of observations into a partitioned space requires that an estimate of some functional which depends on an unknown distribution over a probability space $(\Omega, F\ , P)$.

We cannot observe the true distribution $(P)$ directly. To but we can bound the uncertainty about the distribution, Vapnik uses a set of hypotheses as to whether the data being classified $(x_i \in \Omega)$ belongs to a partition or not. The set of labels $\{l_1, l_2, \cdots, l_K\}$ partitions the entire sample space $\left(\bigcup l_i \equiv \Omega\right)$. [15]

Vapnik in his work has proposed a series of bounds on or measures of the risk of making a classification error. These bounds are best understood in the context of a simple 2 dimensional example (Fig. 12). In this example, we are trying to find a line on a plane that best separates two sets of observations. The observations have 2 elements $(x_{1i}, x_{2i})$ or observed characteristics associated with each point and a label $\{-1, +1\}$. So the goal of the classifying machine is to map the observed data $\left(x_i = \begin{bmatrix} x_{1i} & x_{2i} \end{bmatrix}^T\right)$ into a labeled set associated with the data $\left(y_i \in \{-1, +1\}\right)$. [15]

Training this type of learning machine is a supervised process. It is assumed that we have access to a set of observations where we know for certain the desired classification or set that the data points belong to. The data is separated into a random set of points used to train set and a verification set.

The independence of the data used for training is the use of sampling of the available data set. It is the random sampling that also reduces the machine's dependence on correlation between the observations that form the observed set of data. This reduced dependence on the correlation will in general make the machine's performance more robust assuming that the observations are of an independent and identically distributed random process.

Since we have no control over the data sets we deal with, it is important to be able to perform the classification when the data are ill-conditioned or there is not enough information in the data to precisely specify the solution. In these situations, the problem is transformed into an alternative representation in a Hilbert space. An additional term is introduced to condition or regularize the data.

The use of a Hilbert space representation allows for the introduction of a kernel function. In this study, the kernel function has the following form

$$k(\underline{u}, \underline{v}) = \left(0.6\underline{u}^T \underline{v}\right)^3 \qquad (21)$$

The SVM training algorithm seeks to find those values of $\underline{w}$ that minimizes the risk function.

$$J = \gamma k\left(\underline{w}^T, \underline{w}\right) + k\left(\left(\underline{y} - \underline{w}^T \underline{x}\right)^T, \left(\underline{y} - \underline{w}^T \underline{x}\right)\right) \qquad (22)$$

This function seeks to balance the complexity of the classifier (norm of the weights $\left(k\left(\underline{w}^T, \underline{w}\right)\right)$) and accuracy in separating the points into partitions $\left(k\left(\underline{\varepsilon}^T, \underline{\varepsilon}\right)\right)$. Note that the perpendicular distance, $\varepsilon_i = \underline{y}_i - \underline{w}^T \underline{x}_i$, is the error of the linear classification function at a given data vector, $\underline{x}_i$, of the training data set.

The conditioning term, $\gamma$, controls the degree of regularization. From another point of view, the conditioning term controls the level of complexity of the support vector machine. Ensuring that the machine is not over fitted to the training data set and reducing the machine's ability to classify future values. The lower the number of non-zero weights the smaller the sum resulting from the norm. Thus minimizing the number of support of vectors need to span the data space.

The verification set is used to assess the machines ability to classify points in general (not in the training set). It is important to remember this distinction. It is the general performance that reflects the machine's ability to classify future as of yet unknown observations. The separation of the data into these groups ensures statistical independence between the training and risk assessment process.

## VI. Experimentation

The sound recording we used are from a training video

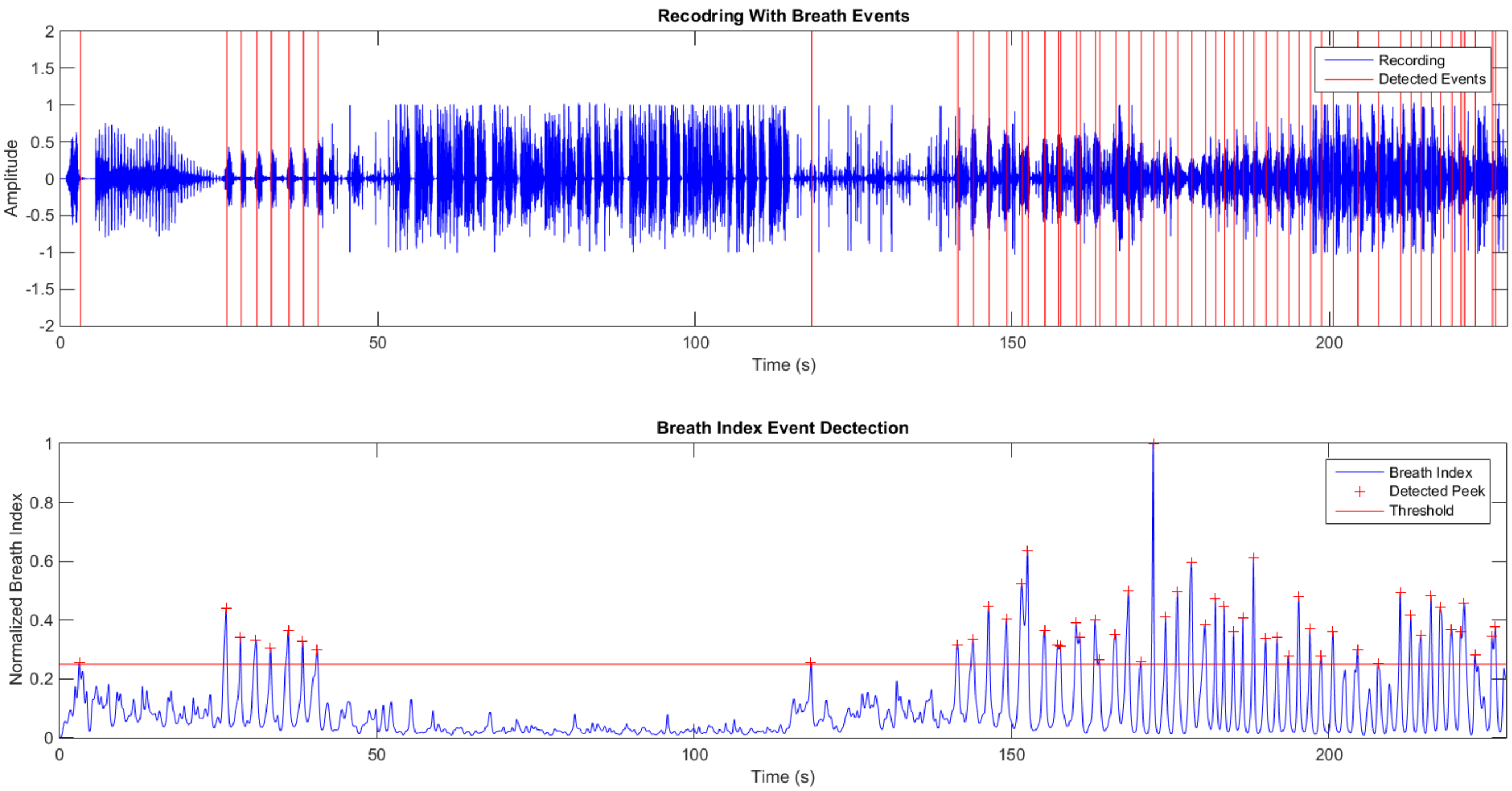

Fig. 13, The top plot shows the audio track as separated from the training video. The red lines represent the detected regulator inspiration events. Not all events were detected. As this is a function of the threshold chosen and exemplars used to form the statistics for the breath index calculations. The bottom figure represents the normalized breath index (B/max(B)). The red line represents the threshold used (0.25) and the red crosses the peeks detected by the MATLAB peek detection function.

(“FIREGROUND Fire Entrapment - Conserving SCBA Air”) [9] intended to educate firefighters on what to do when their low air alarm goes off. The tape begins with some music followed by the training breathing using the SCBA equipment. He then takes the mask off and lectures. After the lecture he puts the mask back on and then proceeds to use a stair climber to simulate doing the labor a fireman does. The low alarm goes off and the trainer proceeds to review air conservation procedures with the mask on.

In a near real time application data would be processed as each frame is completed. However, in this study, the entire file is read in and processed into a cepstrogram. Then using a sliding window of the same width as the templates. The window is advanced or slide to the right by one cepstral column. The breathiness index computation is then repeated for the data falling inside the window

.

Fig. 13 summarizes the detection performance. It should be noted that the performances is dependent on the threshold level and the data used in the template.

The LPC approach provides a different result. The figure shows several regions where the algorithm has improperly identified as a breathing events. The use of additional requirements including the duration length and power levels result in the data about a detected breathing event to be available after the event with many false starts. The prediction errors are presented in Fig. 17. The radial plot assigns event an angular measurement. The magnitude of the error for each sampled point is plotted along the angular vector.

The SVM approach allows us to classify the LPC gain data in real time with no false starts. Comparing the errors between the two techniques, Fig. 17 to Fig. 18, it is clear that the SVM approach does a better job of identifying events than the LPC only approach. Both LPC and SVM approach show a improvement over the pattern matching approach (Fig. 16).

Fig. 15 and Fig. 14 show the errors in determining duration of each breathing event. The next stage of processing involved computing the breathing rates in hertz between each detection event as summarized in TABLE I. The breathing periods are estimated by subtracting consecutive breathing event times. These time differences are then inverted to get the breathing rate in Hertz.

TABLE II tabulates the measured inhalation lengths or durations using the LPC SVM approach. These were determined by subtracting the time that an inhalation events starts from the time it ends.

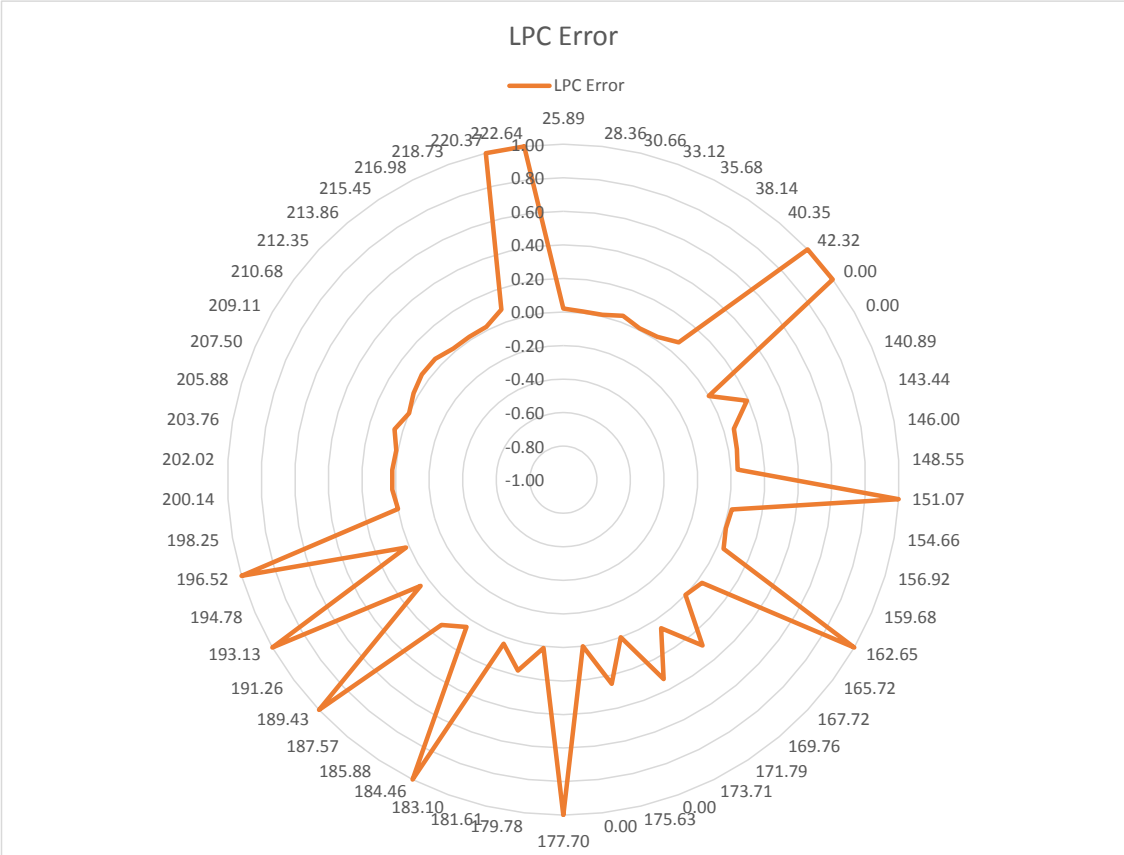

Fig. 17 Radial error magnitude plot for the identifying the start of a breathing event using the LPC algorithm and the secondary constraints. The spikes with magnitude of 1 represent events not detected or anticipated by the filter. The shorter spikes represent differences between the reference data and the filter responses.

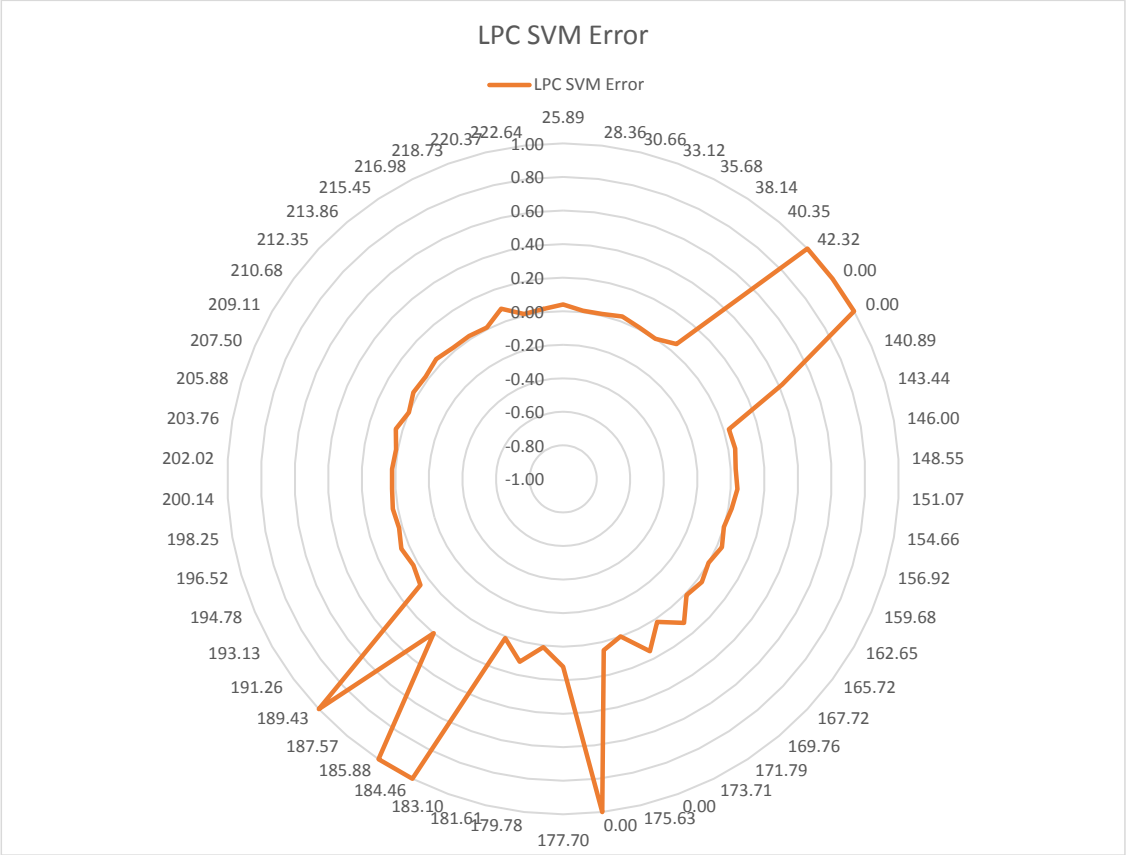

Fig. 18 Radial error magnitude plot for the identifying the start of a breathing event using the LPC algorithm and a Support Vector Machine. The values at magnitude 1 represent events not detected or anticipated. The other much smaller spikes represent the smaller errors than using the LPC and secondary constraints.

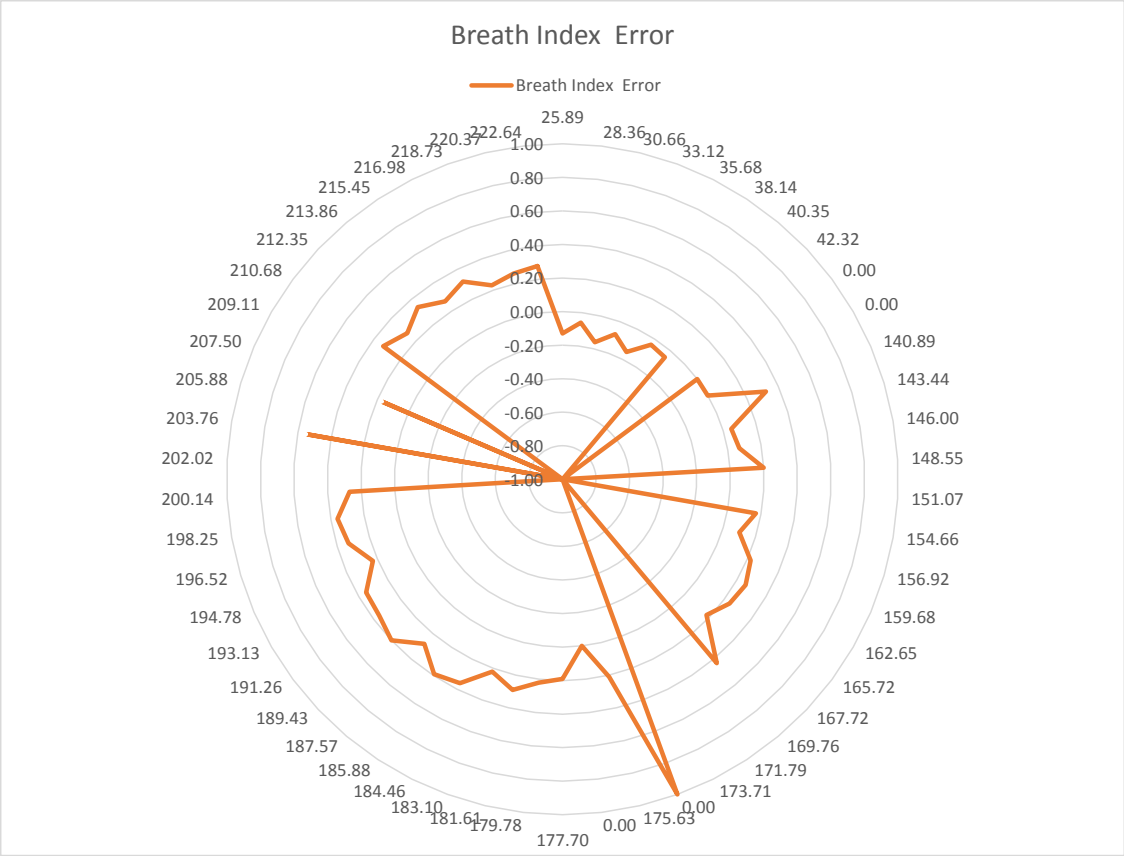

Fig. 16 Radial error magnitude plot for the identifying the start of a breathing event using the using the Breathing Index and Support Vector Machine. The spikes with magnitude of 1 represent events not detected or anticipated by the filter. The shorter spikes represent differences between the reference data and the filter responses.

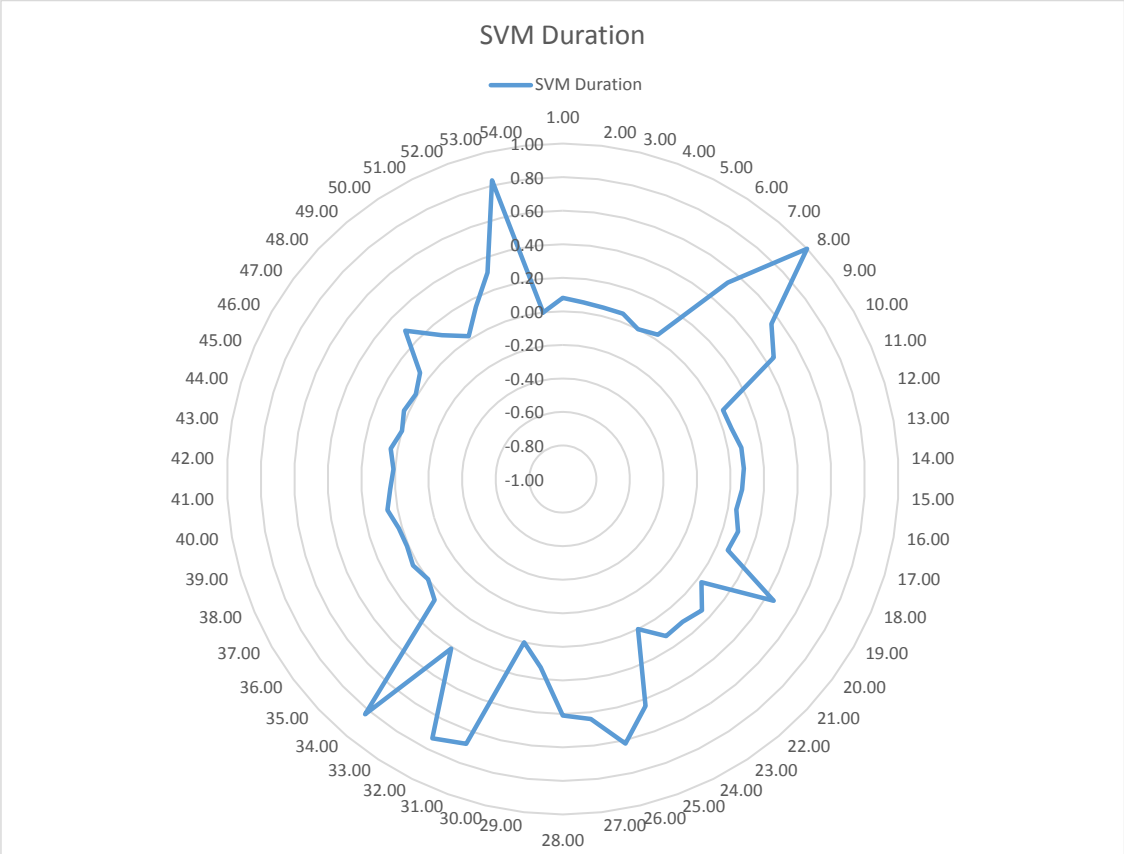

Fig. 15 Radial error magnitude plot for the identifying the duration of a breath using the LPC algorithm and SVM. The spikes with magnitude of 1 represent events not detected or anticipated by the filter. The shorter spikes represent differences between the reference data and the filter responses.

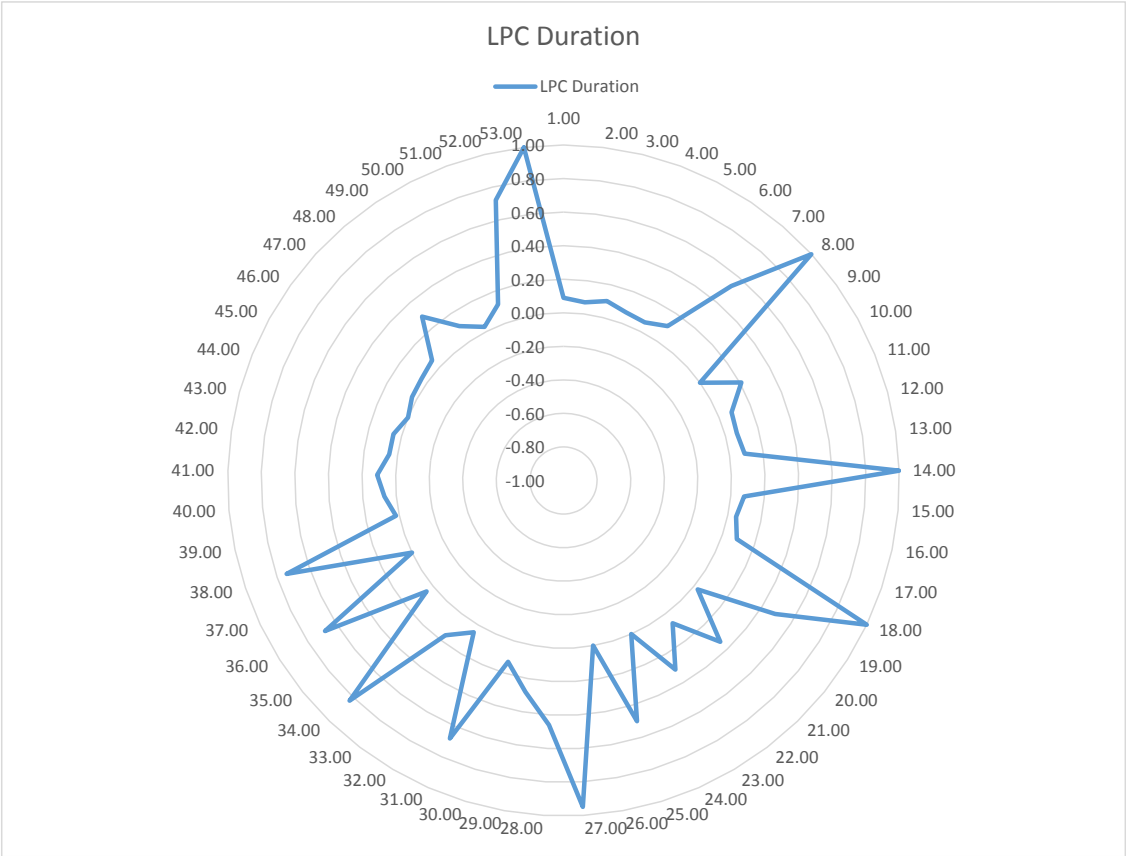

Fig. 14 Radial error magnitude plot for the identifying the duration of a breath using the LPC algorithm and the secondary constraints. The spikes with magnitude of 1 represent events not detected or anticipated by the filter. The shorter spikes represent differences between the reference data and the filter responses.

The medical references measure the number of breaths in a minute as the standard. An estimate the breathing rate of an adult at rest of approximately 15 to 20 breaths per minute. A histogram of the data converted to breaths per minute (Fig. 20), shows the start of a separation of the main data into two modes. The modes represent both the instructors breathing at the start of the video and also after he has performed some physical activity at the end. The large valued outliers represent false detections where the time between detected values are extremely short. The values around zero probably represent the intervals where the instructor is talking but is not wearing the mask.

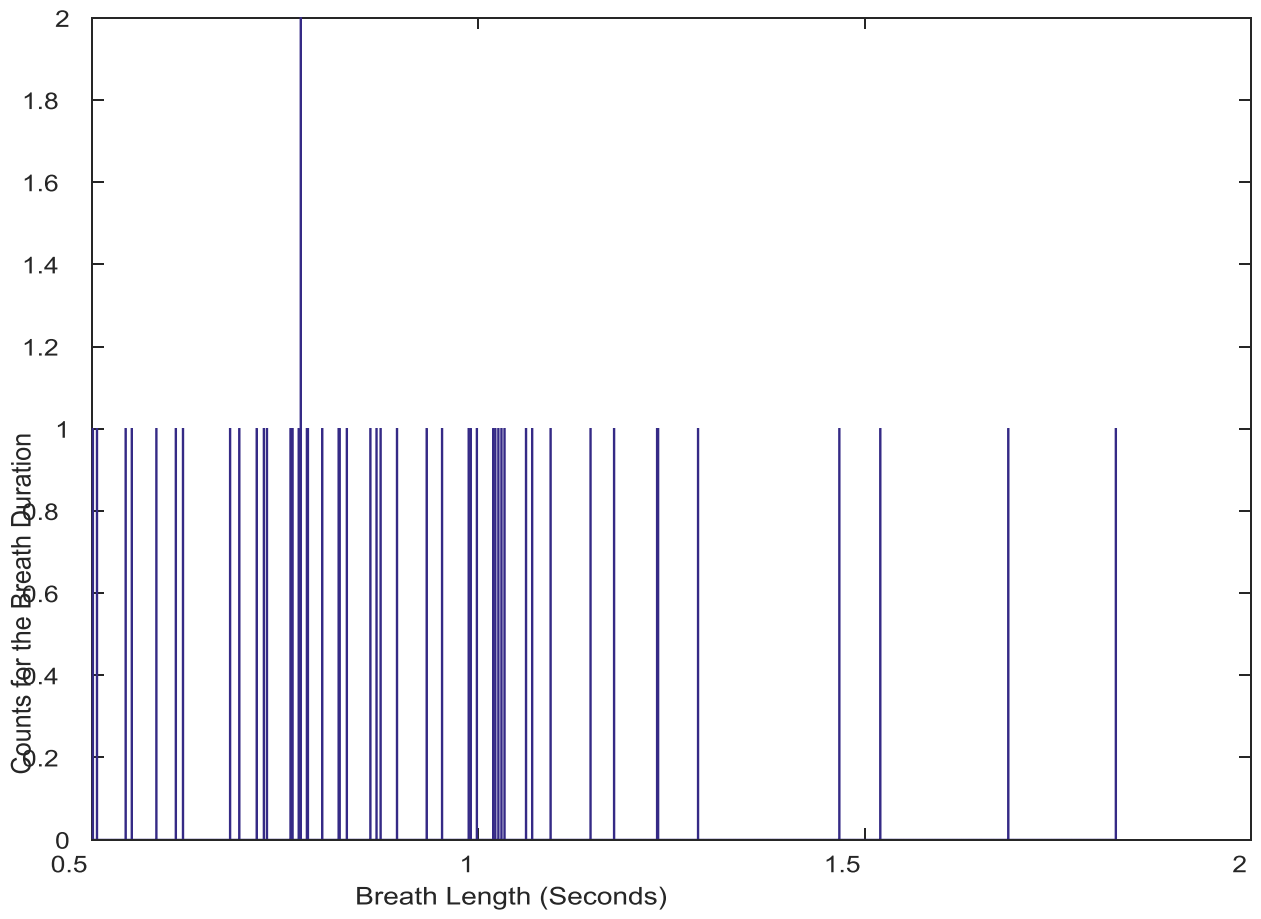

Fig. 19, Histogram of the breathing durations. Note the outliers. These either represent false detections where the time between detected values are extremely short. The values around zero probably represent the intervals where the instructor is talking but is not wearing the mask.

TABLE I

MEASURED BREATHING RATES

| Event Time (sec) | Rate (bpm) | Event Time (sec) | Rate (bpm) | Event Time (sec) | Rate (bpm) |
|---|---|---|---|---|---|
| 25.93 | 2.63 | 162.65 | 20.00 | 198.28 | 34.36 |
| 28.35 | 24.79 | 165.69 | 19.72 | 200.12 | 32.60 |
| 30.65 | 26.10 | 167.73 | 29.47 | 200.12 | 32.60 |
| 33.09 | 24.63 | 169.64 | 31.42 | 202.00 | 31.94 |
| 35.69 | 23.06 | 171.77 | 28.18 | 203.77 | 34.03 |
| 38.14 | 24.54 | 173.56 | 33.45 | 205.84 | 28.92 |
| 40.30 | 27.68 | 175.68 | 28.30 | 207.50 | 36.08 |
| 127.80 | 0.69 | 177.13 | 41.31 | 209.08 | 38.07 |
| 133.63 | 10.28 | 177.82 | 87.80 | 210.66 | 38.10 |
| 141.31 | 7.82 | 179.79 | 30.39 | 212.31 | 36.31 |
| 143.41 | 28.53 | 181.49 | 35.20 | 213.84 | 39.16 |
| 145.96 | 23.50 | 183.11 | 37.08 | 215.43 | 37.66 |
| 148.52 | 23.49 | 187.77 | 12.89 | 216.97 | 39.10 |
| 151.03 | 23.82 | 191.32 | 16.86 | 218.65 | 35.62 |
| 154.64 | 16.63 | 193.16 | 32.70 | 220.36 | 35.06 |
| 156.92 | 26.38 | 194.83 | 36.01 | 222.62 | 26.57 |
| 159.65 | 21.97 | 196.54 | 35.05 | | |

TABLE II

MEASURED BREATH LENGTHS

| Event Time (sec) | Duration (sec) | Event Time (sec) | Rate (bpm) | Event Time (sec) | Rate (bpm) |
|---|---|---|---|---|---|
| 25.93 | 0.99 | 162.65 | 1.52 | 198.28 | 0.77 |
| 28.35 | 0.80 | 165.69 | 1.28 | 200.12 | 0.82 |
| 30.65 | 1.03 | 167.73 | 1.02 | 200.12 | 0.72 |
| 33.09 | 0.77 | 169.64 | 1.07 | 202.00 | 0.95 |
| 35.69 | 1.02 | 171.77 | 0.77 | 203.77 | 0.76 |
| 38.14 | 0.76 | 173.56 | 1.09 | 205.84 | 0.76 |
| 40.30 | 1.00 | 175.68 | 0.54 | 207.50 | 0.78 |
| 127.80 | 1.23 | 177.13 | 0.62 | 209.08 | 0.83 |
| 133.63 | 0.55 | 177.82 | 0.51 | 210.66 | 0.82 |
| 141.31 | 0.71 | 179.79 | 1.15 | 212.31 | 0.87 |
| 143.41 | 0.99 | 181.49 | 0.90 | 213.84 | 0.87 |
| 145.96 | 0.93 | 183.11 | 0.61 | 215.43 | 1.03 |
| 148.52 | 1.18 | 187.77 | 0.58 | 216.97 | 0.86 |
| 151.03 | 1.83 | 191.32 | 0.78 | 218.65 | 1.03 |
| 154.64 | 1.06 | 193.16 | 0.68 | 220.36 | 0.50 |
| 156.92 | 1.23 | 194.83 | 0.69 | 222.62 | 0.77 |
| 159.65 | 1.47 | 196.54 | 0.73 | | |

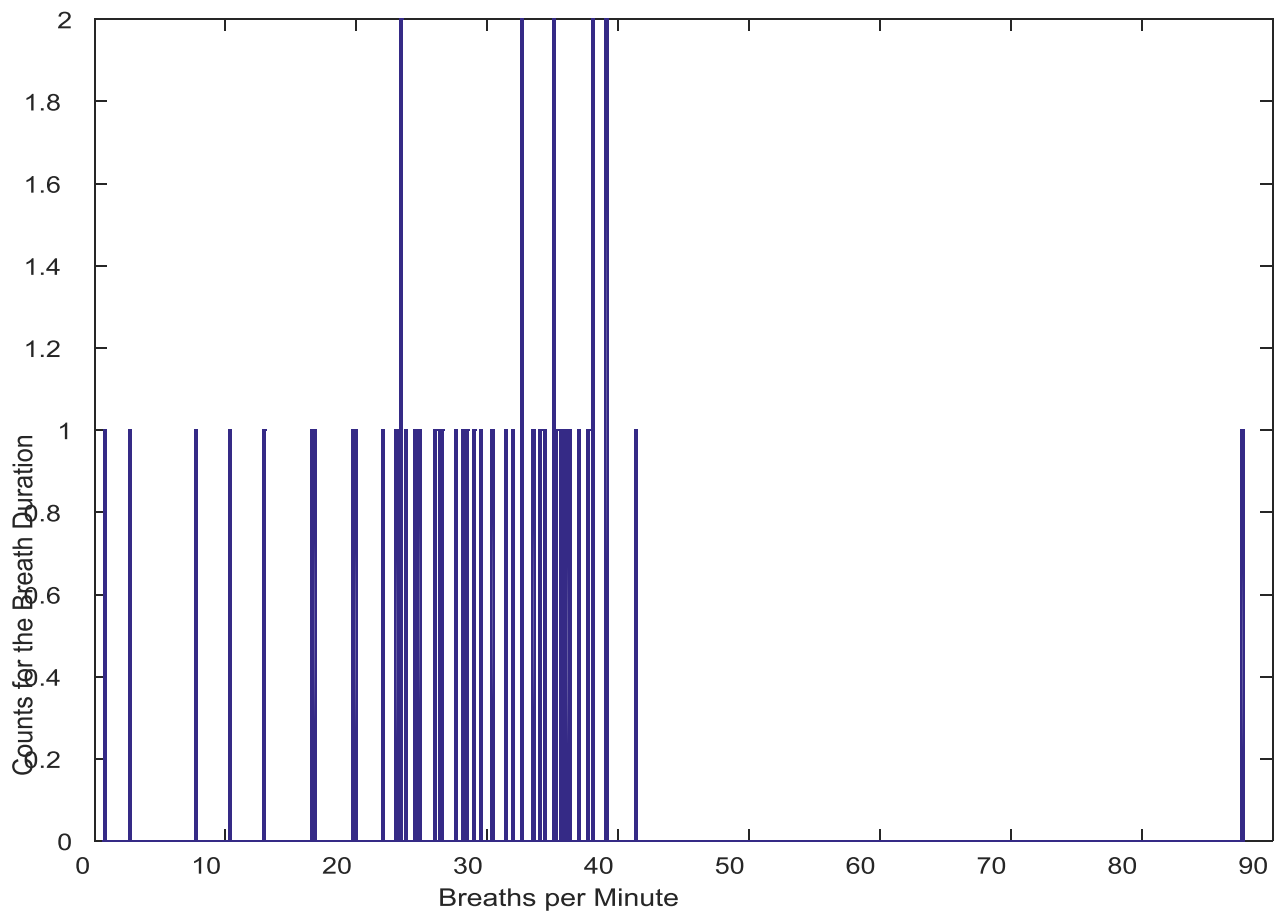

Fig. 20, Histogram of the breathing rates. Note the outliers. These either represent false detections where the time between detected values are extremely short. The values around zero probably represent the intervals where the instructor is talking but is not wearing the mask.

## VII. Conclusion

The methods studied indicate that an approach like this is promising but is heavily dependent on having a high energy signature event being monitored. It is also sensitive to the threshold values being used. An alternative approach using wavelets might be able to yield a simpler algorithm for detect breathing events.

We will continue to refine the process by developing a machine learning technique/algorithm that will focus on classifying the breathing rate as normal or an outlier and also to be able to determine at rest and under stress levels of breathing.